\documentclass[conference]{IEEEtran}
\IEEEoverridecommandlockouts

\usepackage{cite}
\usepackage{amsmath,amssymb}
\usepackage{graphicx}
\usepackage{adjustbox}
\usepackage[caption=false,font=footnotesize]{subfig}
\usepackage{booktabs}
\usepackage[hyphens]{url}
\usepackage{xcolor}

\usepackage[colorinlistoftodos,textsize=footnotesize]{todonotes}

\graphicspath{{images/}}

\begin{document}

\title{ZoionNET: Low-Power Animal Intrusion Detection in the South
Indian Hills Using an Acoustic Deep-Learning Classifier for a
Cellular Remote-Sensing System\thanks{This work received an Honorary
Mention in the 2024 IEEE ComSoc Student
Competition~\cite{comsoc2024winners}.}}

\author{
\IEEEauthorblockN{Prasanna Vijay\IEEEauthorrefmark{1},
Adarsh Gupta\IEEEauthorrefmark{2},
Vansh Panwar\IEEEauthorrefmark{2},
Vatsalya Singh\IEEEauthorrefmark{2},\\
Shubham Singhal\IEEEauthorrefmark{2},
Anubhav Saha\IEEEauthorrefmark{2},
Teena Sharma\IEEEauthorrefmark{3}}
\IEEEauthorblockA{%
\IEEEauthorrefmark{1}Sri Sairam Engineering College (Anna University),
Chennai, India\\
\IEEEauthorrefmark{2}Indian Institute of Technology Guwahati,
Assam, India\\
\IEEEauthorrefmark{3}Mehta Family School of Data Science and Artificial
Intelligence,\\
Indian Institute of Technology Guwahati, Assam, India}
}

\maketitle

\begin{abstract}
Crop raiding and village intrusion by elephants and large felids impose a
recurring cost on communities along forest boundaries in southern India,
and existing camera-based monitoring has proven difficult to sustain on
the power budgets available at such sites. We present ZoionNET, a
battery-powered sensing node that detects animal intrusion acoustically
rather than visually. Audio from an onboard PDM microphone is sampled at
8~kHz, converted to mel-spectrograms on-device, and classified on an Arm
Ethos-U55 microNPU by two int8-quantised YOLOv8 models arranged as a two-state
machine: a lightweight always-on model runs in a dormant state and gates a
four-class species classifier through exponential-moving-average
thresholds, so the heavier model runs only when an intrusion is likely. A
digital vibration sensor supplies a complementary seismic channel.
Inferences leave the node as JSON over a cellular Notecard, reaching a web
dashboard and SMS alerts to forest rangers without any local gateway. The
two models reach reported accuracies of 0.974 and 0.946 on held-out
audio, and a complete node costs \$133.57 in parts. We report the system
as built and deployed, and identify explicitly what has not yet been
measured --- most notably the energy behaviour that motivates the
two-state design.
\end{abstract}

\begin{IEEEkeywords}
acoustic event detection, edge machine learning, human--wildlife conflict,
microNPU, cellular IoT, wildlife monitoring
\end{IEEEkeywords}

\section{Introduction}
\label{sec:intro}

The mid-altitude hill ranges of Southern India encompass extensive forest divisions that adjoin cultivated land along a long and largely unfenced boundary. This perimeter operates less as a barrier than as an intersection: wild fauna cross it routinely, with adjacent agricultural communities absorbing the primary consequences. Elephants serve as the flagship species in these fragmented landscapes, where habitat degradation has rendered seasonal movements and crop-foraging patterns increasingly unpredictable in recent years~\cite{shaji2019kans}. Every season,
farmers around Hosur lose standing crops and occasionally their lives.
The cost runs in both directions --- elephants that come into repeated
contact with human activity die from shooting, road accidents and
electrocution, so keeping them away from settlements is itself a
conservation measure.

Large felids compound the problem. In early January 2021, a leopard
entered a village near the Melagiri forest region of Tamil Nadu, killing
a large number of stray dogs and livestock and attacking several people;
in the same period, a large elephant herd damaged crops in the
surrounding farmland. Locally, the severity of the episode was
attributed to the absence of any early-warning or monitoring system that
could detect animal movement before it reached the village.

Early warning is therefore the intervention that matters, and the
constraint that governs it is power. A monitoring node on this boundary
has no mains supply and, at the distances involved, no practical
maintenance schedule. Camera-based approaches, which dominate automated
wildlife monitoring, are a poor fit: continuous video capture and
inference exhaust a battery long before the deployment interval is over,
and outside protected areas camera traps fill with livestock and people
within days (Section~\ref{sec:background}).

ZoionNET responds by changing the sensing modality rather than the
hardware budget. Sound carries around vegetation and over distance,
costs far less to sample than video, and --- once reduced to a
mel-spectrogram --- is classifiable by the same convolutional
architectures that serve image tasks. Pairing an acoustic channel with a
seismic one, and classifying the audio on a microNPU that idles between
inferences, yields a node that can plausibly sit on a battery at a forest
edge and speak to a ranger's phone over the cellular network that already
covers the area.

Elephants, the dominant source of conflict along this boundary, are the
prototype's primary target. The classifier also covers monkeys and large
felids; together with elephants, these are the animals the project's
supporting NGO identifies as the source of roughly 90\% of intrusion
incidents at the site. Its felid class was trained on cheetah
vocalisations, so it demonstrates that the pipeline can learn a big-cat
class rather than serving as a validated detector for the leopards found
locally; retraining it on leopard recordings is part of the remaining
work.

This paper reports the system as built and field-trialled. Its
contributions are:

\begin{itemize}
\item An acoustic and seismic sensing node built on a microNPU-class edge
      platform, deployable unattended at a forest--village boundary.
\item A two-state inference design in which a lightweight always-on
      classifier gates a heavier species classifier through
      exponential-moving-average thresholds, so that the expensive model
      runs only when an intrusion is likely.
\item An end-to-end cellular delivery path from edge inference to a web
      dashboard and SMS alerts, requiring no local gateway
      infrastructure.
\item A complete node costing \$133.57 in parts, reported against a
      published single-board-computer system addressing the same
      problem.
\end{itemize}

We also state plainly what the deployment has not yet established. The
energy argument that motivates the two-state design is architectural
rather than measured, and Section~\ref{sec:limitations} sets out that gap
alongside the others.

\section{Background}
\label{sec:background}

\subsection{Monitoring outside protected areas}

Much of India's wildlife lives outside its protected-area network, which
covers only about 5\% of the country; roughly 80\% of the Asian
elephant's range in India lies beyond it~\cite{bhandari2026wire}. That is
where conflict happens: on farmed, grazed and inhabited land along the
forest edge. Local conservation organisations work with forest
departments to reduce it~\cite{shaji2019kans}, but this is also where
standard monitoring breaks down. Camera traps, the default tool, fill
with livestock and people within days, the most informative positions are
the ones most likely to be disturbed, and servicing units means long
journeys on foot; work in these landscapes is also often short of funding
and specialist equipment~\cite{bhandari2026wire}.

For an early-warning node, these conditions set two requirements: it must
run unattended on its own power for long periods, and it must report only
the animals that matter rather than everything that moves past it. Recent
published systems take a camera-based route to the second requirement:
Ronoh et al.~\cite{ronoh2022hwc} pair a passive infrared trigger with a
camera and run YOLO object detection on a Raspberry Pi, delivering SMS
alerts over GSM. The pipeline is close to what is proposed here, but the
sensing modality and the compute platform place it in a different power
class --- a comparison we return to in Section~\ref{sec:results}.

\subsection{Why acoustic and seismic sensing}

Both alternative modalities have a track record. Passive acoustic
monitoring is established for elephants in particular: Bjorck et
al.~\cite{bjorck2019elephant} built a large crowdsourced corpus of
forest-elephant recordings and showed that learned models can detect and
segment calls from continuous audio. On the seismic side, Wijayaraja et
al. demonstrated that elephant footfalls produce ground vibration
detectable at range, first building a geophone--sensor interface suitable
for embedded systems~\cite{wijayaraja2024seismic} and subsequently
classifying those events with machine learning at ranges up to roughly
150~m~\cite{wijayaraja2025seismic}.

What both modalities share is a far lower acquisition cost than video.
Audio at 8~kHz is four orders of magnitude smaller per second than even
modest video, and the sensor itself draws negligible current. Once a
short window of audio is converted to a mel-spectrogram, the
classification problem becomes a small image problem, which is convenient
because that is precisely the regime in which convolutional networks are
efficient. Hershey et al.~\cite{hershey2017cnn} established that
image-derived CNN architectures transfer well to spectrogram-based audio
event classification at scale, working over the AudioSet
ontology~\cite{gemmeke2017audioset}.

\subsection{Enabling hardware}

The remaining obstacle is that a convolutional network still has to run
somewhere. Microcontroller-class parts have historically been too slow
for this, and application processors too power-hungry. Arm's Ethos-U55
microNPU is designed for exactly this gap: an accelerator small enough to
sit beside a Cortex-M core, which Arm reports as delivering up to a 90\%
energy reduction for machine-learning workloads compared with previous
Cortex-M generations~\cite{arm2024ethosu55}. Realising that benefit requires
the network to be quantised to integer arithmetic, following the
now-standard int8 quantisation-aware training approach of Jacob et
al.~\cite{jacob2018quantization}.

Together these make it feasible to run a modern detector architecture
on a battery-powered node --- which is what the remainder of this paper
describes.

\section{System Architecture}
\label{sec:arch}

\begin{figure}[t]
\centering
\includegraphics[width=\columnwidth]{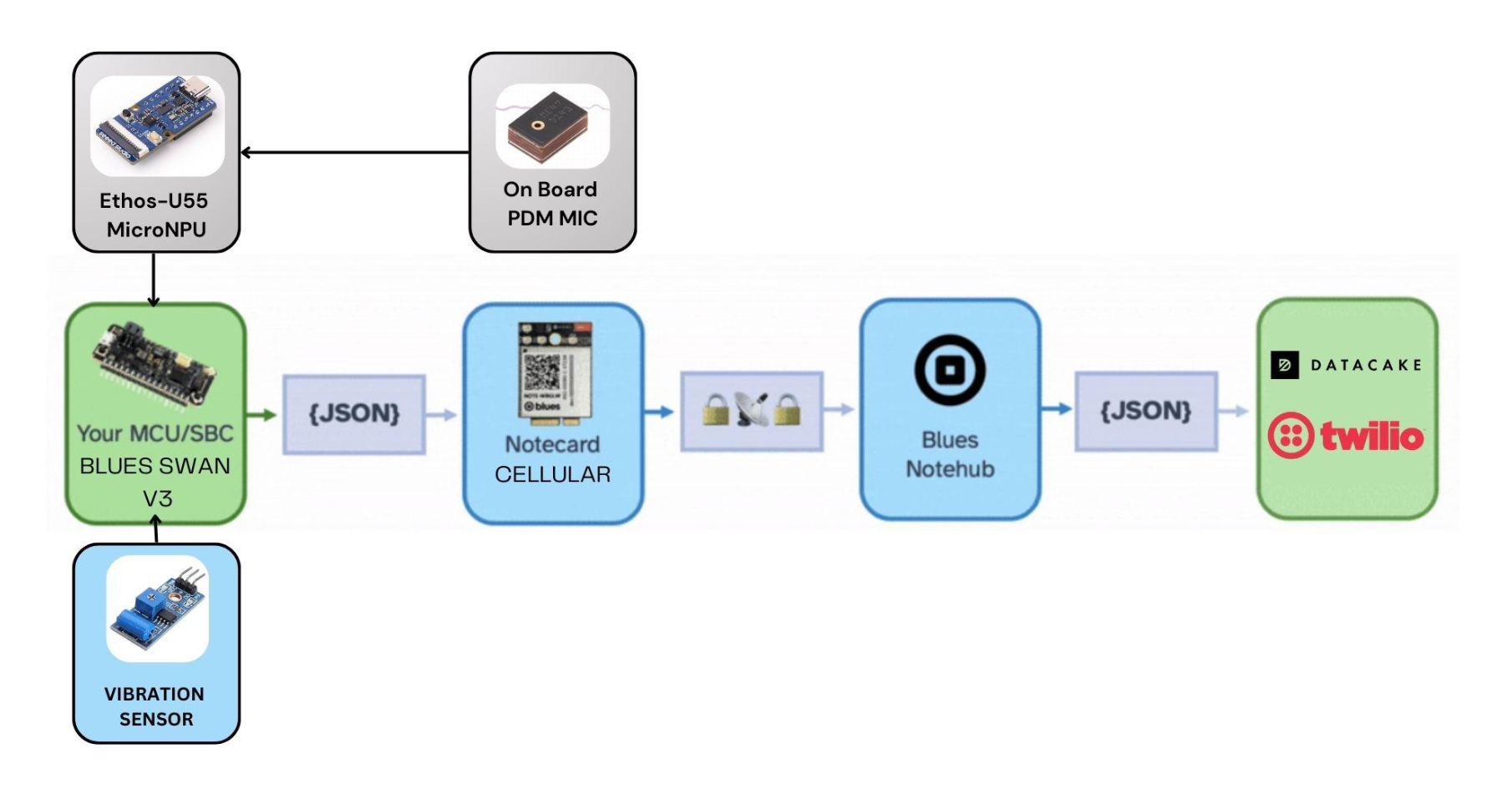}
\caption{ZoionNET system architecture. Audio from the onboard PDM
microphone is classified on an Arm Ethos-U55 microNPU, whose inferences
pass over I\textsuperscript{2}C to a Blues Swan V3 STM32
microcontroller; the vibration sensor connects to the Swan directly. The
Swan emits JSON payloads over cellular to Blues Notehub for routing to
Datacake and Twilio.}
\label{fig:arch}
\end{figure}

Figure~\ref{fig:arch} shows the node end to end. It divides into three
parts: the hardware platform, an edge inference path that turns sound
into a class label, and a connectivity path that carries that label to
the people who act on it.

\subsection{Hardware platform}

The node is assembled from commodity modules rather than custom silicon,
which keeps both the cost and the reproduction effort low.

\textbf{Sensing.} Audio is captured by the PDM microphone integrated on
the Seeed Grove Vision AI V2 module~\cite{seeed2024grovevision}. Pulse
density modulation encodes the waveform as a high-frequency one-bit
stream rather than an analogue level, which suits a digital front end and
avoids a separate ADC. An SW-420 vibration switch module supplies the
seismic channel: a normally closed vibration switch whose state an
onboard LM393 comparator turns into a clean digital output. It is
intended to register the ground-borne signature of heavy animals such as
elephants --- the modality studied in detail by Wijayaraja et
al.~\cite{wijayaraja2024seismic}.

\textbf{Inference.} The Grove Vision AI V2 is built around a Himax
WiseEye2 HX6538 with a dual-core Arm Cortex-M55 and an integrated Arm
Ethos-U55 microNPU~\cite{seeed2024grovevision}. Arm offers the Ethos-U55
in configurations of 32 to 256 MACs per cycle, rated at 64 to 512~GOP/s
at 1~GHz~\cite{arm2024ethosu55}. The HX6538 integrates the 64-MAC
configuration clocked at up to 400~MHz~\cite{himax2023hx6538}, which
places its peak throughput at roughly 51~GOP/s (64 MACs, two operations
each, at 400~MHz). That is a ceiling rather than a delivered rate: Arm
quotes MAC utilisation of up to 85\% on popular networks.

\textbf{Control and connectivity.} A Blues Swan V3, an STM32-based
development board, acts as the node's controller: it receives inferences
from the microNPU, applies the payload thresholds, and drives the radio.
The radio itself is a Blues Notecard cellular System-on-Module carrying
its own Cortex-M4, GPS, accelerometer and temperature sensor, seated on a
NoteCarrier-A carrier board that supplies the antenna and the physical
interface to the Swan. Blues ships cellular Notecards with ten years of
service and 500~MB of prepaid data included, with no SIM or monthly
subscription~\cite{blues2024notecard}.

\textbf{Power and storage.} A single lithium-ion cell powers the node,
with a 32~GB microSD card for local retention. Capacities are given in
Section~\ref{sec:results}.

\subsection{Sensing and edge inference path}

The microphone delivers a continuous time-domain waveform to the
HX6538, sampled at 8~kHz. A sliding window segments that stream, and
each window is converted on-device into a mel-spectrogram --- a
frequency-domain image of the segment, with time on one axis and
frequency on the other. Section~\ref{sec:pipeline} gives the transform
and the models that consume it.

Classification runs entirely on the node. When a window is classified,
the HX6538 emits an inference --- a class label together with a
confidence score --- to the Swan V3 over
I\textsuperscript{2}C. Firmware on the microcontroller evaluates that
stream against the thresholds described in
Section~\ref{sec:pipeline} and decides whether the event warrants a
transmission. No alert leaves the node unless it does, which is what keeps
the cellular duty cycle low.

The seismic channel takes a simpler path that bypasses the microNPU. The
Swan V3 reads the SW-420's digital output directly in one-second windows,
counting a sample roughly every 0.1~ms while the output is high, so the
count reflects how long the switch was triggered within the window. The
count is scaled linearly to a 0--1023 \texttt{vibration\_value},
saturating at a count of 150: stronger, longer shaking gives a higher
value. The firmware reports values of 600 and above as an elephant and
250--599 as a large cat (labelled \emph{cheetah}, the felid class used
throughout the prototype), sending a \texttt{sensors.qo} note and an SMS
alert in either case. A reading is also uploaded at a fixed interval
regardless of value.

\subsection{Connectivity and cloud path}

The Notecard performs a device-to-cloud data pump to Blues Notehub, a
proxy and routing service. Synchronisation is periodic, or immediate
when an intrusion is detected. Payloads are JSON, chosen so that the radio moves as few bytes
as possible per event. Beyond sensor readings and inference results, the
Notecard exchanges session housekeeping with Notehub: opening and closing
sync connections, brown-out and hard-reset notifications, and session
boundaries.

Sensor and inference data travel in a \texttt{sensors.qo} note:

{\small
\begin{verbatim}
"file": "sensors.qo",
"body": {
  "animal": "elephant",
  "description": "Vibration threshold
      exceeded, data of an elephant
      sent.",
  "vibration_value": 968
}
\end{verbatim}
}

Notehub acknowledges the session, accepts the payload, stores it, and
routes it onward over HTTPS to two consumers. Datacake~\cite{datacake2024}
receives \texttt{sensors.qo} via an HTTP webhook pointed at its API
endpoint, and provides visualisation, device-status monitoring and
long-term retention. Twilio Programmable
Messaging~\cite{twilio2024messaging} receives a separate
\texttt{twilio.qo} note, which Notehub converts into an HTTP POST that
triggers an SMS to forest rangers and to the supporting NGO:

{\small
\begin{verbatim}
"file": "twilio.qo",
"body": {
  "customFrom": "+125661XXXXX",
  "customMessage": "Alert - Animal
      Intrusion Occurred - Elephant",
  "customTo": "+91XXXXXXXXXX"
}
\end{verbatim}
}

Routing both notes through Notehub rather than having the node contact
each service directly means the device holds no third-party credentials
and the alert destinations can be changed without touching deployed
firmware.

\section{Acoustic Classification Pipeline}
\label{sec:pipeline}

\begin{figure*}[t]
\centering
\subfloat[Monkey]{\includegraphics[width=0.72\textwidth]{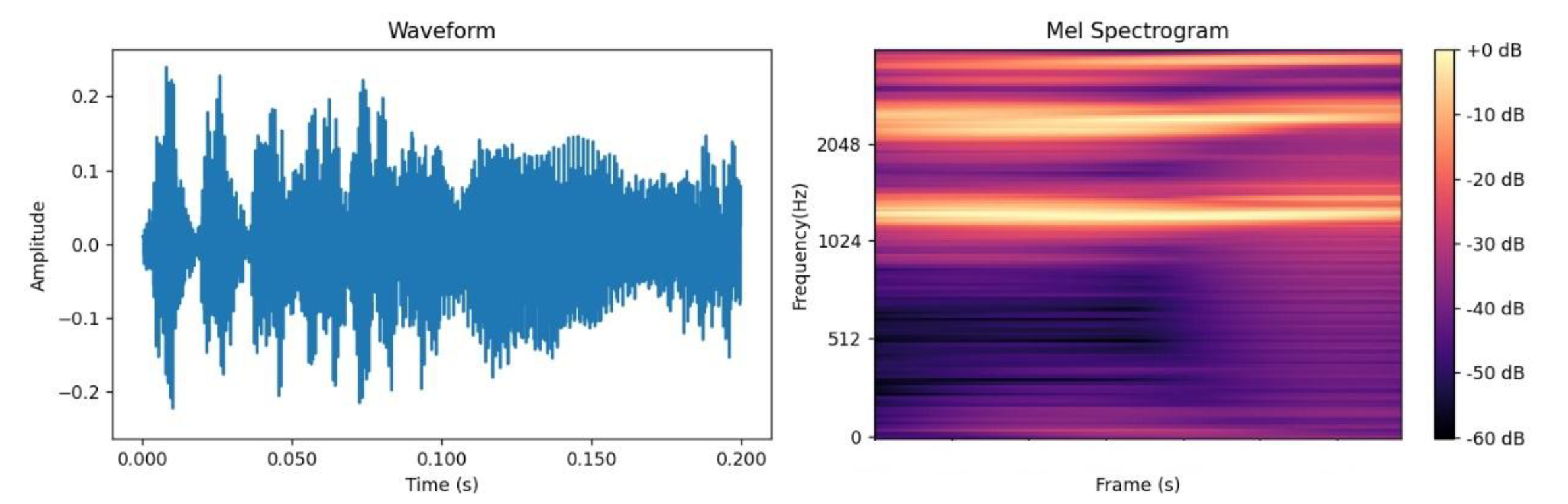}}\\
\subfloat[Elephant]{\includegraphics[width=0.72\textwidth]{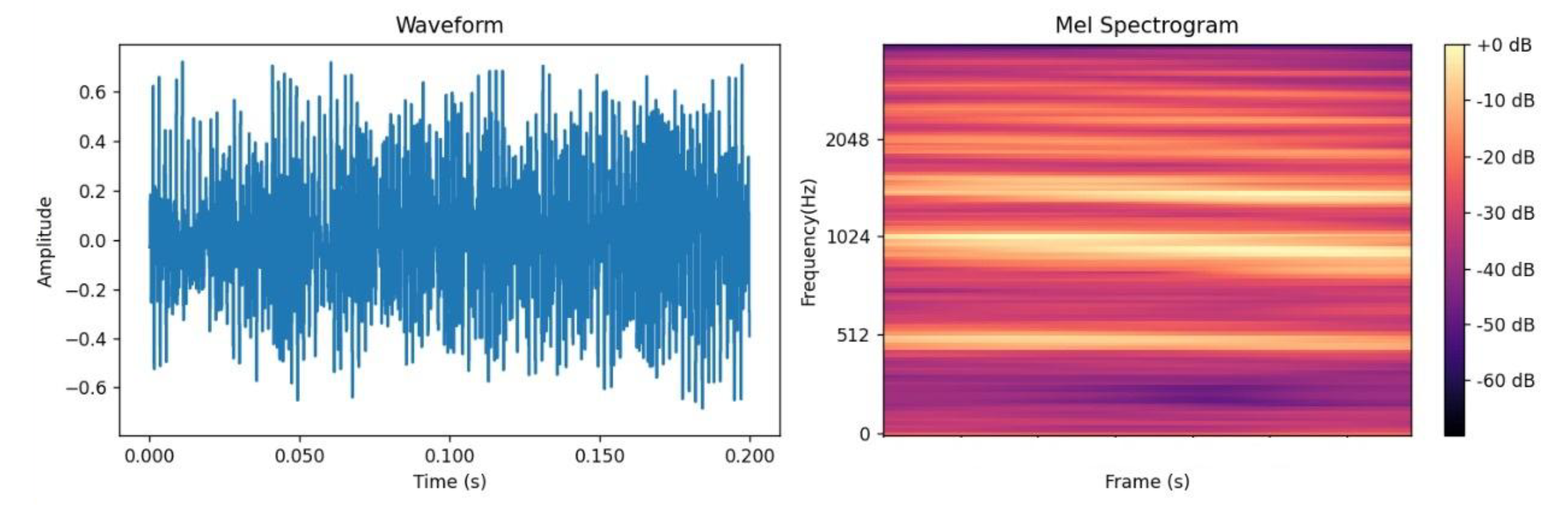}}\\
\subfloat[Cheetah]{\includegraphics[width=0.72\textwidth]{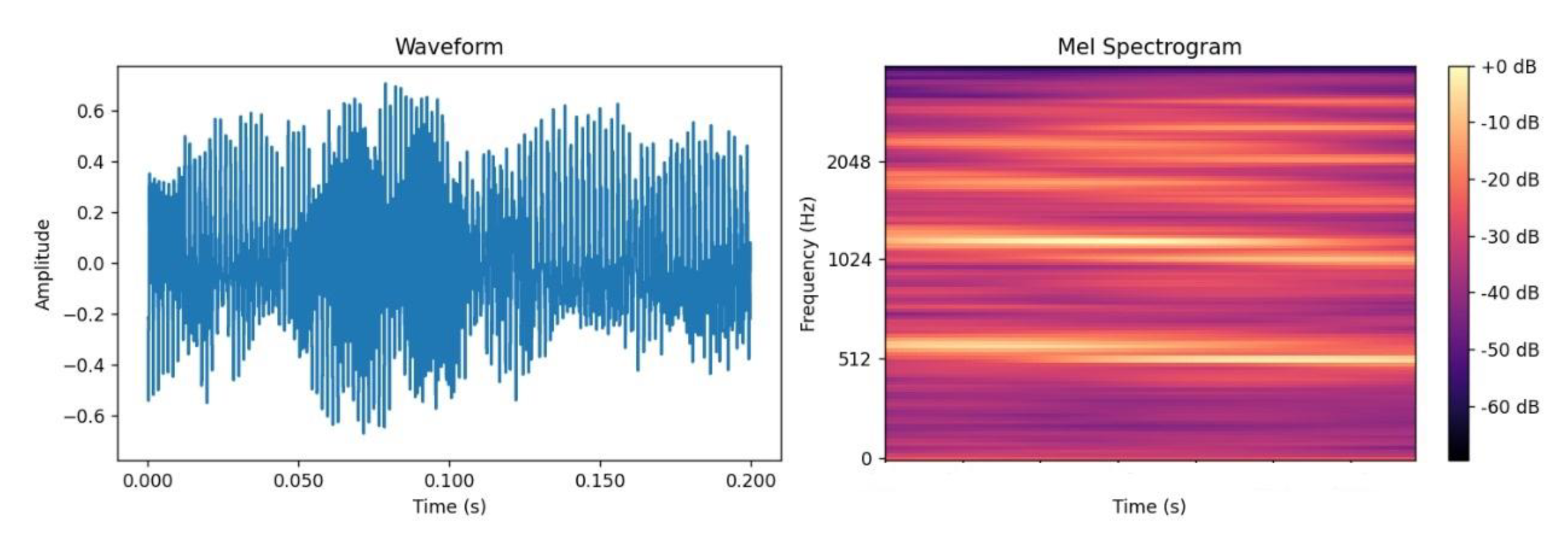}}
\caption{Time-domain waveform (left of each pair) and the corresponding
mel-spectrogram (right) for a 0.2~s segment of each target class,
sampled at 8~kHz. The classes separate visibly in the frequency domain:
the monkey segment concentrates energy above 1~kHz with a quiet band
below 512~Hz, while the elephant and cheetah segments carry strong
harmonic structure through the 512~Hz--2~kHz range.}
\label{fig:spectrograms}
\end{figure*}

\subsection{Acquisition, short-time Fourier transform and mel warping}

A raw waveform is a poor input to a convolutional classifier: the
information that distinguishes a trumpet from a bird call lives in how
spectral energy is distributed and how that distribution evolves, not in
the sample values themselves. The node therefore converts audio to a
time--frequency image before classifying it.

Audio arrives from the PDM microphone at 8~kHz. A sliding window of
0.2~s advances every 0.1~s, so consecutive windows overlap by half their
length; the overlap means a vocalisation straddling a window boundary
still appears intact in one of the two. Each window is transformed by the
short-time Fourier transform~\cite{allen1977stft}, which applies an
analysis window $w$ at hop $H$ and takes the discrete Fourier transform
of each windowed segment:

\begin{equation}
X(m,k) = \sum_{n=0}^{N-1} x[n]\,w[n-mH]\,
         e^{-j 2\pi k n / N}
\label{eq:stft}
\end{equation}

Here $m$ indexes the frame and $k$ the frequency bin, so $|X(m,k)|$ is a
two-dimensional array --- time along one axis, frequency along the other.
The sliding window thus captures both short-term fluctuation within a
call and longer-term structure across successive frames.

The linear frequency axis of~\eqref{eq:stft} is then warped onto the mel
scale, an empirical mapping from frequency to perceived
pitch~\cite{stevens1937mel} conventionally approximated as

\begin{equation}
m(f) = 2595 \log_{10}\!\left(1 + \frac{f}{700}\right)
\label{eq:mel}
\end{equation}

and implemented as a bank of overlapping triangular filters. The warping
compresses high frequencies relative to low ones, which concentrates
resolution where animal vocalisations carry most of their discriminative
structure and reduces the input dimensionality the network must handle.
This representation has been the standard front end for audio recognition
since Davis and Mermelstein~\cite{davis1980mfcc}, and transfers directly
to convolutional architectures at scale~\cite{hershey2017cnn}.

The result is a 128-band mel-spectrogram over a 1600-sample (0.2~s)
window. Figure~\ref{fig:spectrograms} shows the waveform and
mel-spectrogram for one segment of each class. Critically, the same
transform runs on the HX6538's Cortex-M55 cores at inference time as was
used to build the training set, so the images the deployed model sees
match those it was fitted to.

\subsection{Dataset}

Training audio was aggregated from public sources, principally YouTube
and Freesound~\cite{font2013freesound}, selected to approximate the
acoustic conditions of the deployment site as closely as available
material allowed. Four classes were used: background noise including
birds, cheetah, elephant, and monkey.

Bird audio was deliberately over-represented. Cheetah vocalisations
include a chirp that correlates strongly with bird calls in the
mel-spectrogram domain, and without additional negative examples the
detection model confused the two.

\subsection{Models and fine-tuning}

Two models were fine-tuned from pre-trained YOLOv8
checkpoints~\cite{jocher2023yolov8}, the current iteration of the
single-stage YOLO family~\cite{redmon2016yolo}. Both are classifiers
operating on mel-spectrogram images. Their backbone is a convolutional
feature extractor followed by a neck performing multi-scale feature
aggregation, then global average pooling and fully connected layers
terminating in a softmax.

The \emph{detection model} is binary. Its negative class holds birds and
background noise, its positive class the three target species. Its role
is not to identify an animal but to answer, cheaply and continuously,
whether anything worth examining is present.

The \emph{classification model} is four-way: background noise and birds,
cheetah, elephant, monkey. It runs only when the detection model has
already indicated activity.

Both were fine-tuned under a 1000-epoch budget with 200-epoch patience,
and both stopped early --- the detection model at approximately 280
epochs and the classification model at approximately 330. Both were then
quantised to int8~\cite{jacob2018quantization} for deployment on the
microNPU, which reduces the memory footprint roughly fourfold against
32-bit floating point and is a precondition for running on the Ethos-U55
at all.

\subsection{Two-state inference with EMA gating}

The two models exist to separate two costs that are usually conflated.
Detecting that \emph{something} is present is cheap; determining
\emph{what} is present is not. Running the four-class model continuously
would spend most of its energy classifying silence.

ZoionNET therefore runs as a two-state machine. In the \textbf{dormant}
state --- the node's normal condition, around the clock --- only the
lightweight detection model runs. When it begins reporting activity the
node transitions to the \textbf{active} state, in which the four-class
classifier identifies the species and payloads are eligible for
transmission. When the classifier's background class comes to dominate
again, the node returns to dormant.

Neither transition is triggered by a single inference. A lightweight
model run continuously will produce occasional false positives, and
waking the full pipeline on each one would erase the saving the design
exists to create. Both transitions are instead driven by an exponential
moving average over the relevant model output:

\begin{equation}
s_t = \alpha\, p_t + (1-\alpha)\, s_{t-1}
\label{eq:ema}
\end{equation}

where $p_t$ is the model's output at step $t$ and $\alpha \in (0,1)$ sets
how quickly the average responds to new evidence. A dormant node
transitions to active when the smoothed detection output crosses a
threshold; an active node returns to dormant when the smoothed
probability of the classifier's background class crosses its own. A
single spurious inference moves $s_t$ by at most $\alpha$, so isolated
false positives are absorbed unless the average is already close to the
threshold; it is sustained evidence that switches the state.

The intent is a node whose expensive component stays idle except when an
intrusion is likely, without sacrificing the sensitivity of the model
that does the identifying.

\section{Implementation and Deployment}
\label{sec:impl}

\subsection{Prototype node}

\begin{figure}[t]
\centering
\includegraphics[width=\columnwidth]{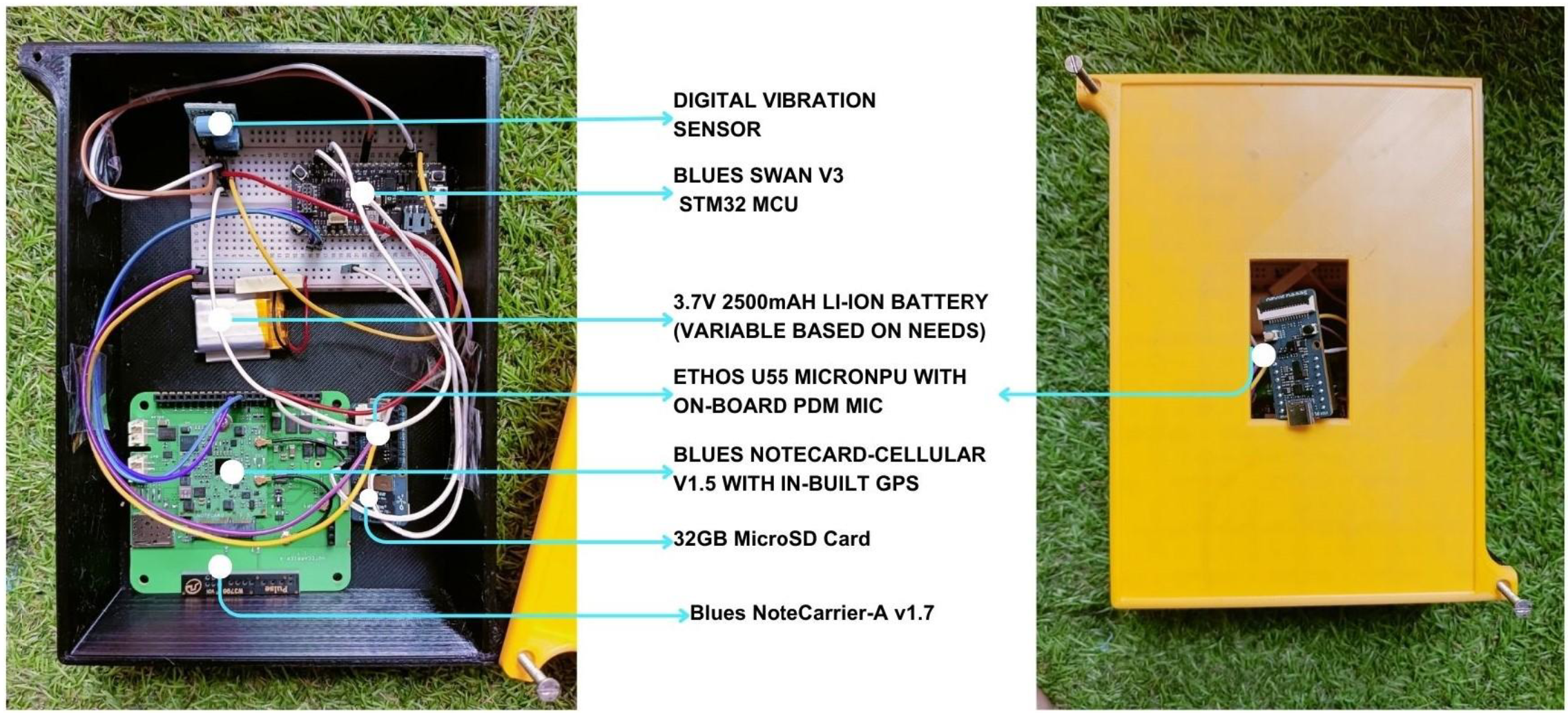}
\caption{Prototype node, opened and closed. Visible inside are the
digital vibration sensor, the Blues Swan V3 STM32 microcontroller, the
lithium-ion cell, the Ethos-U55 microNPU with its onboard PDM
microphone, the Blues Notecard with in-built GPS, a 32~GB microSD card
and the NoteCarrier-A. The aperture in the lid exposes the microphone.
This build carries a single 2500~mAh cell; the bill of materials in
Table~\ref{tab:bom} budgets a 12500~mAh pack sized for unattended field
deployment.}
\label{fig:prototype}
\end{figure}

Figure~\ref{fig:prototype} shows the assembled prototype in a sealed
weatherproof enclosure. The layout is dictated by two constraints that
pull against each other: the microphone must have an acoustic path to the
outside, and the electronics must stay dry through a monsoon. The lid
aperture over the microNPU resolves this, at the cost of being the
enclosure's weakest point.

Battery capacity is deliberately not fixed by the design. The prototype
carries a single 2500~mAh cell, which is convenient for bench work and
short trials; the bill of materials budgets a 12500~mAh pack, which is
what an unattended deployment between service visits requires. Because
the node's consumption has not been characterised (see
Section~\ref{sec:limitations}), the pack size is an estimate rather than
a calculated requirement.

\subsection{Localisation and its fallback}

Each node reports its own position, so that an alert identifies where an
intrusion occurred rather than merely that one did. The Notecard carries
GPS for this purpose.

GPS alone is not dependable at these sites. Canopy cover attenuates the
signal, and the weather conditions under which animals move toward
villages are often the conditions under which a fix is slowest to
acquire. The Notecard therefore falls back to cellular triangulation,
estimating position from the signal characteristics of nearby cell
towers. The estimate is coarser than a GPS fix but requires only the
cellular connectivity the node already depends on to deliver its alerts
--- so any node that can raise an alarm can also say roughly where it is.

\subsection{Field trials}

\begin{figure}[t]
\centering
\includegraphics[width=0.62\columnwidth]{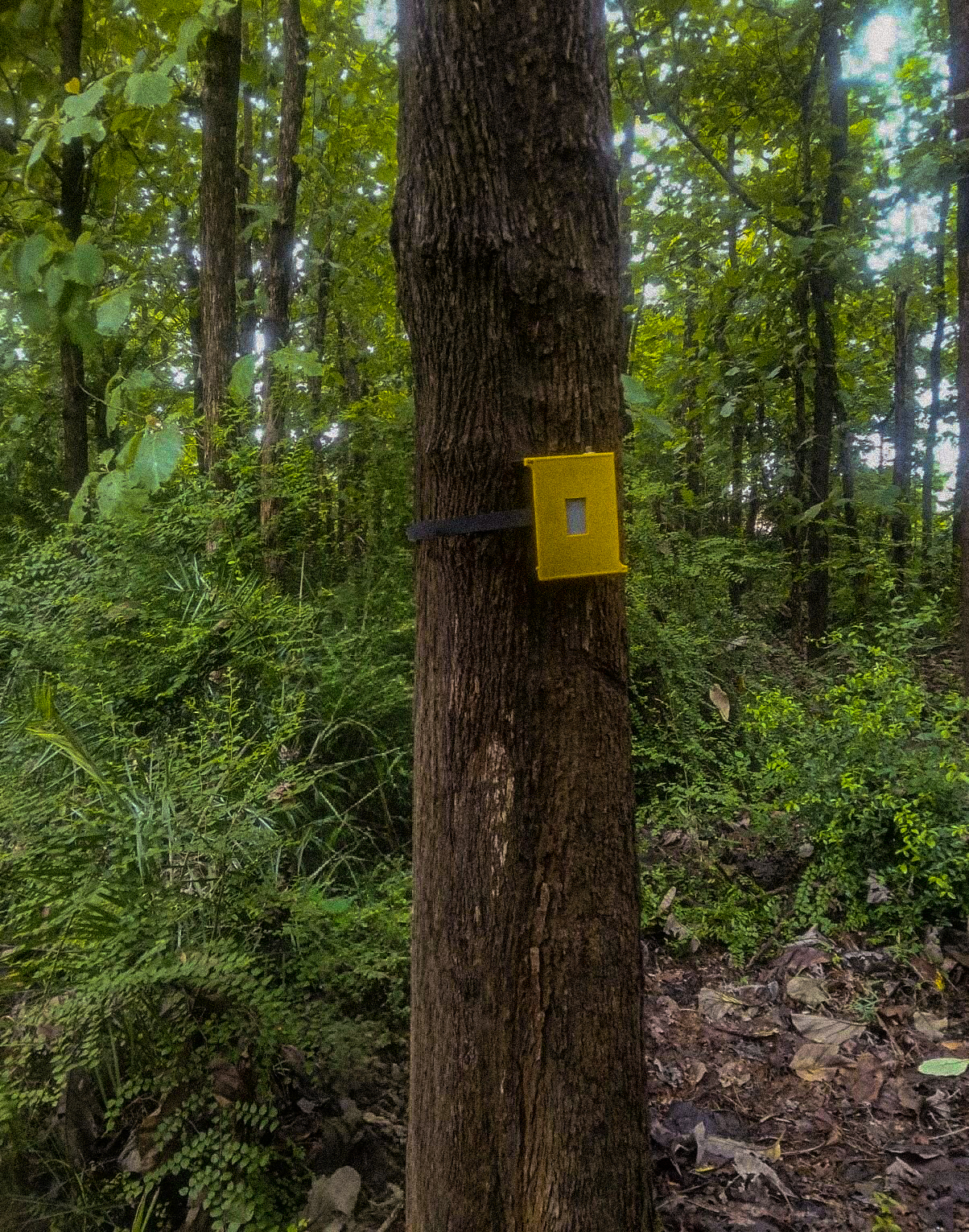}
\caption{A ZoionNET node deployed on a tree at the forest--village
boundary in the Melagiri hills during field trials.}
\label{fig:field}
\end{figure}

Trials to date have evaluated four primary components. Sensor hardware and cellular connectivity were tested in situ within the forest divisions of the Southern Indian Melagiri region, confirming that
the deployment area's cellular coverage is sufficient for the Notecard to
reach Notehub from the boundary itself --- the assumption the whole
communication design rests on. Real-world audio was collected and
annotated at the site. The alert path was integrated end to end, from
on-device inference through to dashboard and SMS. Finally, a pilot node
was deployed in a high-risk area of the forest for validation;
Figure~\ref{fig:field} shows one such unit in place.

\section{Results}
\label{sec:results}

\subsection{Model accuracy}

%
\begin{figure*}[t]
\centering
\subfloat[Training and validation loss]{%
  \includegraphics[width=0.55\textwidth]{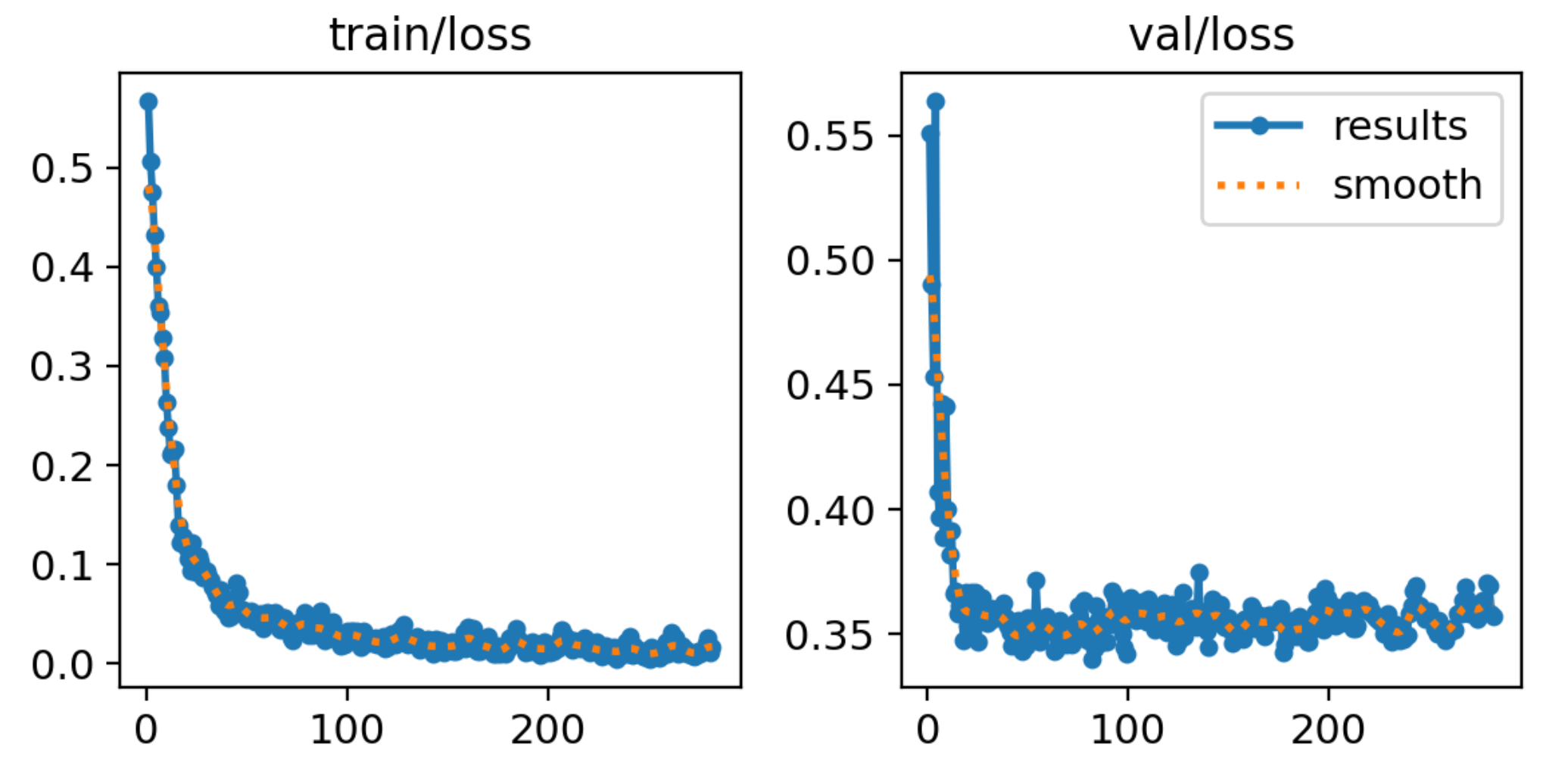}}
\hfil
\subfloat[Top-1 accuracy]{%
  \adjincludegraphics[trim={0 0 {.53\width} 0},clip,width=0.27\textwidth]{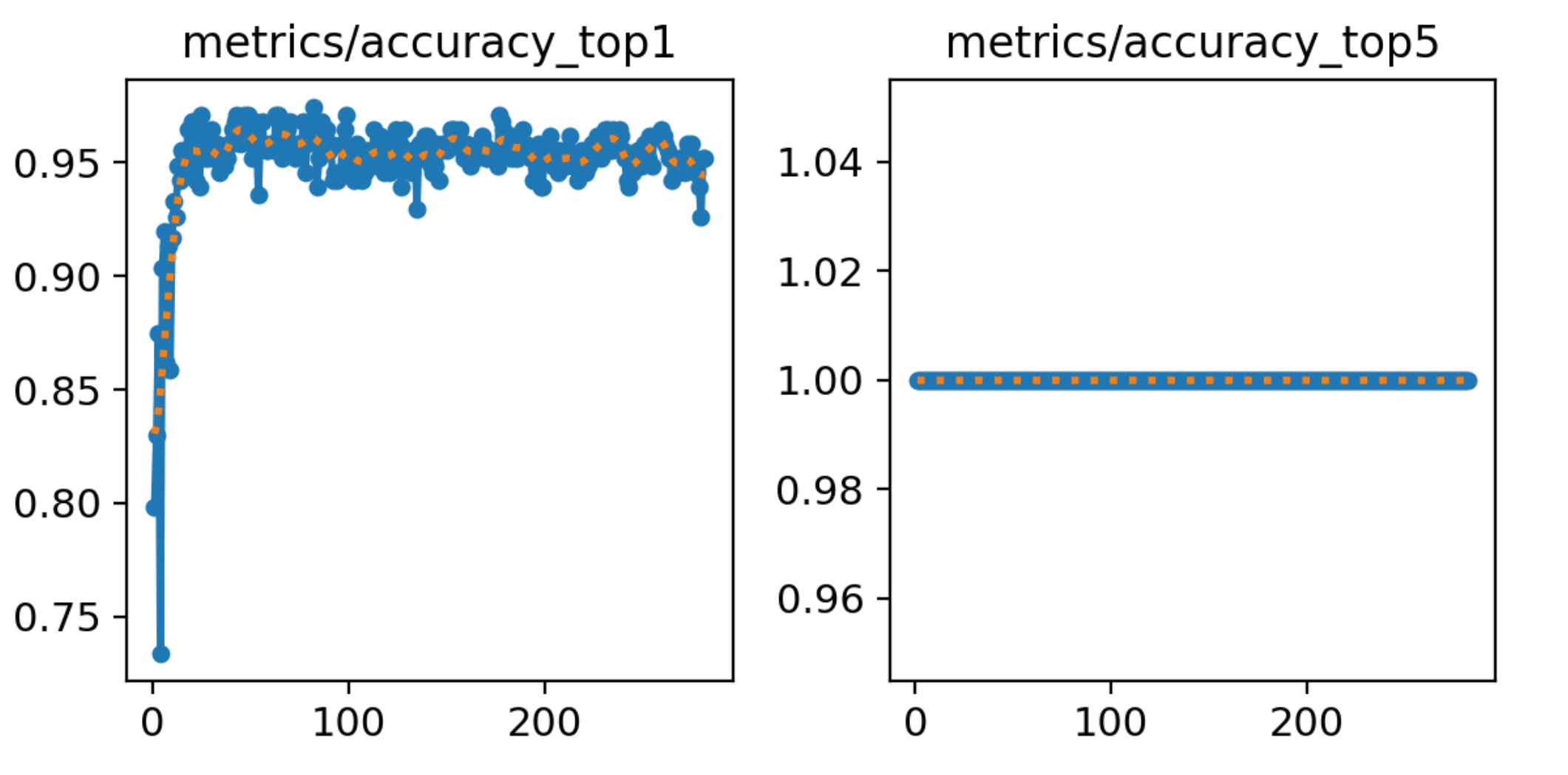}}
\caption{Detection model fine-tuning. Training loss falls sharply within
the first 20 epochs and validation loss plateaus near 0.35; top-1
accuracy settles at approximately 0.95. Early stopping halted training at
approximately 280 epochs under a 1000-epoch budget with 200-epoch
patience.\label{fig:det_curves}}
\end{figure*}

\begin{figure*}[t]
\centering
\subfloat[Training and validation loss]{%
  \includegraphics[width=0.55\textwidth]{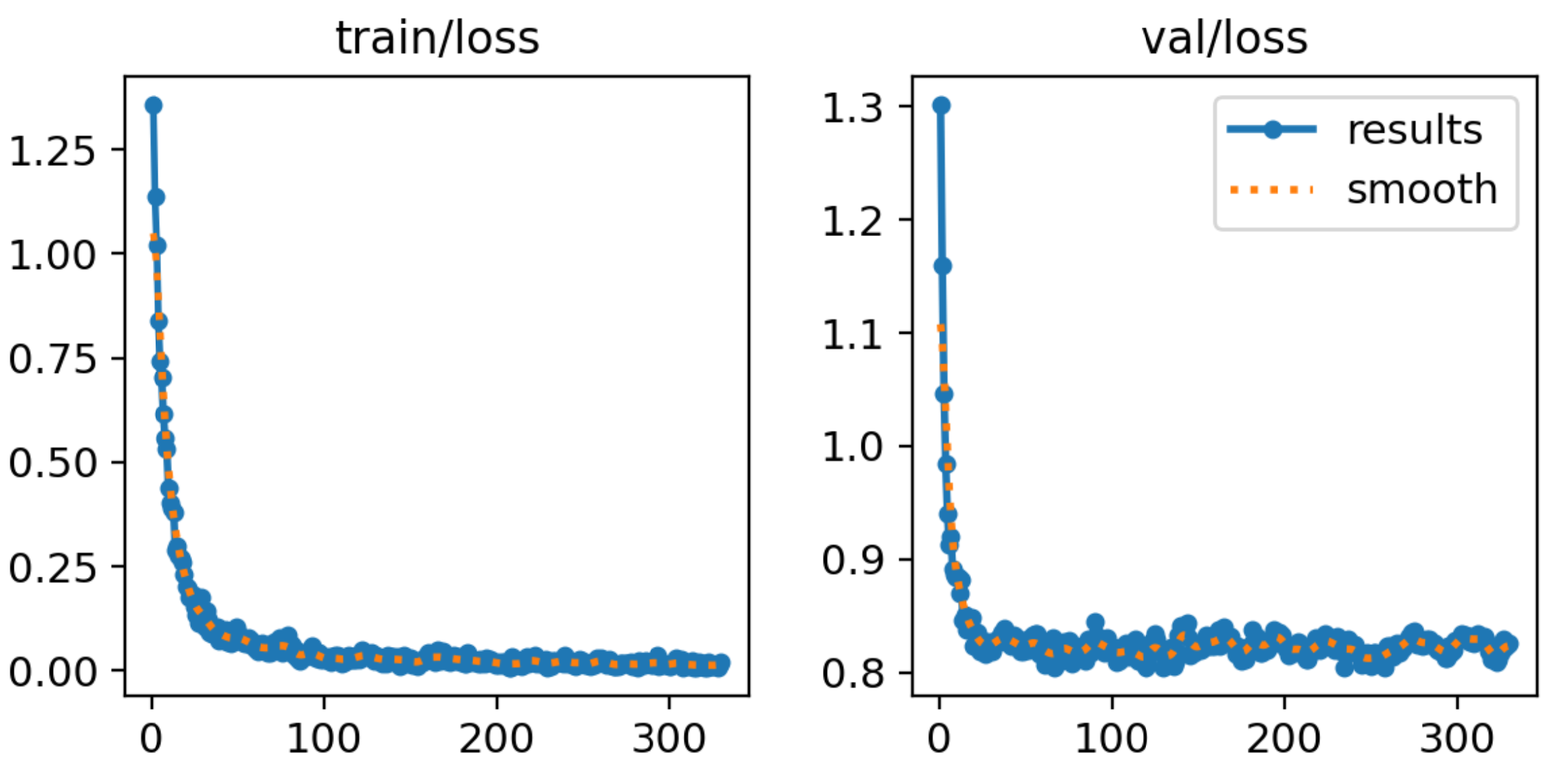}}
\hfil
\subfloat[Top-1 accuracy]{%
  \adjincludegraphics[trim={0 0 {.53\width} 0},clip,width=0.27\textwidth]{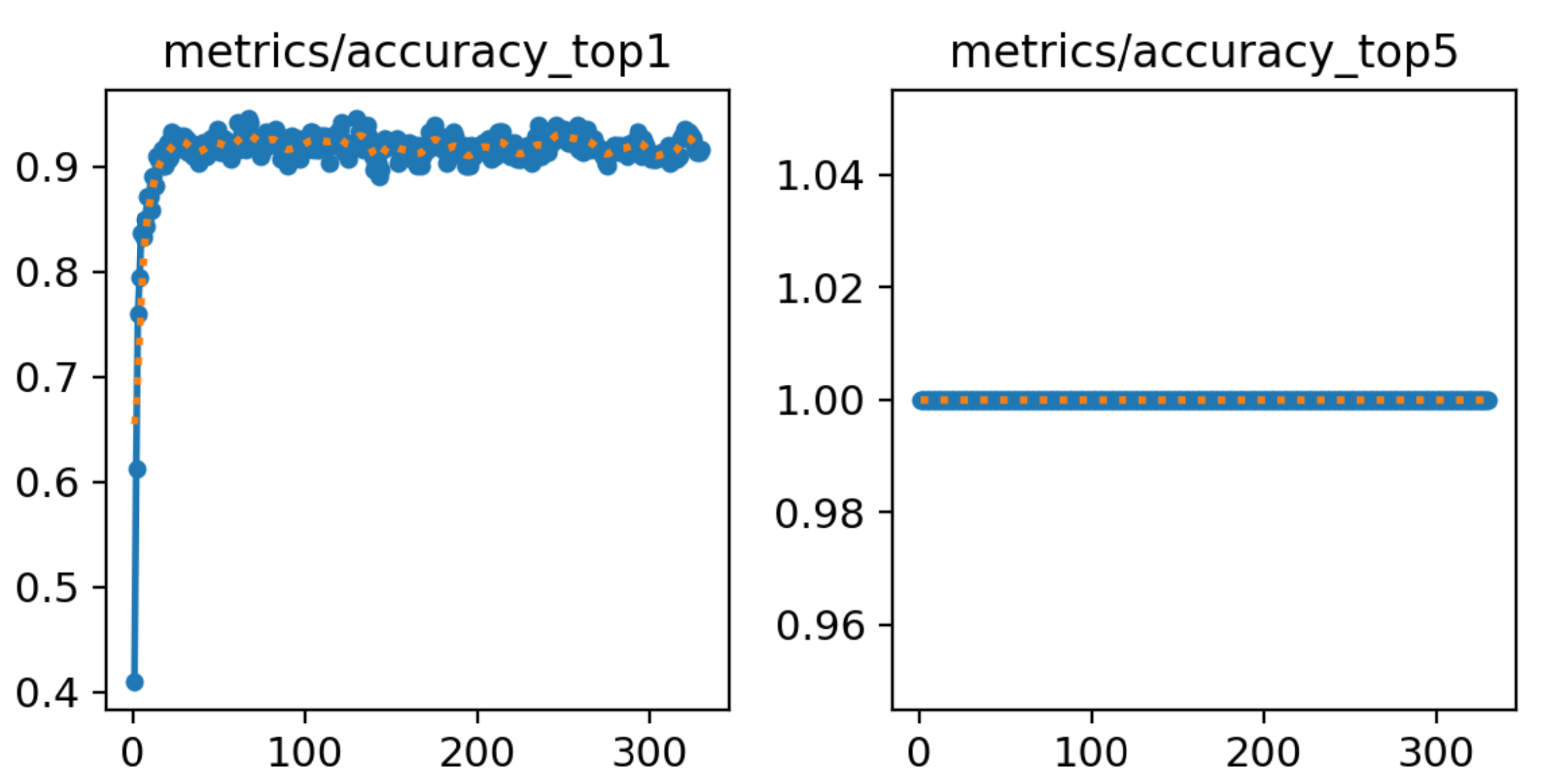}}
\caption{Classification model fine-tuning. Validation loss plateaus near
0.80 and top-1 accuracy settles just above 0.90. Early stopping halted
training at approximately 330 epochs.\label{fig:cls_curves}}
\end{figure*}

The source report gives a final accuracy of 0.97436 for the detection
model and 0.94551 for the classification model.
Figures~\ref{fig:det_curves} and~\ref{fig:cls_curves} show the
corresponding training runs. Both converge quickly --- most of the loss
reduction occurs inside the first 20 epochs --- and both then hold a
stable plateau for the remainder of training, with no divergence between
training and validation loss that would indicate overfitting.

The top-5 accuracy panel present in the original training output is
omitted from both figures. With four classes in one model and two in the
other, top-5 accuracy is 1.00 by construction and carries no information.

\subsection{Cost}

\begin{table}[t]
\centering
\caption{Bill of materials for a single ZoionNET node.}
\label{tab:bom}
\begin{tabular}{@{}llr@{}}
\toprule
Component & Function & Cost \\
\midrule
Blues NoteCarrier-A v1.7 & Carrier board + antenna & \$25.00 \\
Blues Notecard-Cellular & Cellular + GPS module & \$49.00 \\
Blues Swan V3 & STM32 microcontroller & \$24.95 \\
Seeed Grove Vision AI V2 & microNPU + PDM mic & \$15.99 \\
SW-420 vibration sensor & Seismic sensing & \$4.00 \\
Samsung EVO Plus 32\,GB microSD & Storage & \$3.57 \\
3.7\,V 12500\,mAh 18650 Li-ion & Power supply & \$13.00 \\
\midrule
\textbf{Total} & & \textbf{\$133.57} \\
\bottomrule
\end{tabular}
\end{table}

Table~\ref{tab:bom} itemises a complete node at \$133.57. The natural
reference point is the platform the camera-based alternative uses: a
Raspberry Pi~5. Every Pi~5 variant shares the same processor, and the
4\,GB board lists at \$110~\cite{raspberrypi2026pi5}. The two are
therefore close in price, but they do not buy the same thing. The
ZoionNET figure is a complete node: a cellular radio with ten years of
prepaid service, GPS, a battery, storage, a carrier board and both
sensors. The Pi figure is a bare board, which still needs a cellular
modem, a power source sized for an application processor, storage, a
microphone and a vibration sensor before it can do the same job.

This comparison is not hypothetical. Ronoh et al.~\cite{ronoh2022hwc}
build a human--wildlife conflict early-warning system on precisely a
Raspberry Pi with a camera, running YOLO object detection and delivering
SMS alerts --- the same problem, the same alerting model, a different
compute and sensing platform. At comparable board-level cost, what
separates the two approaches is what the money buys and the energy
budget each implies: continuous video inference on an application
processor against gated audio inference on a microNPU.

\subsection{Deployed system output}

\begin{figure}[t]
\centering
\includegraphics[width=\columnwidth]{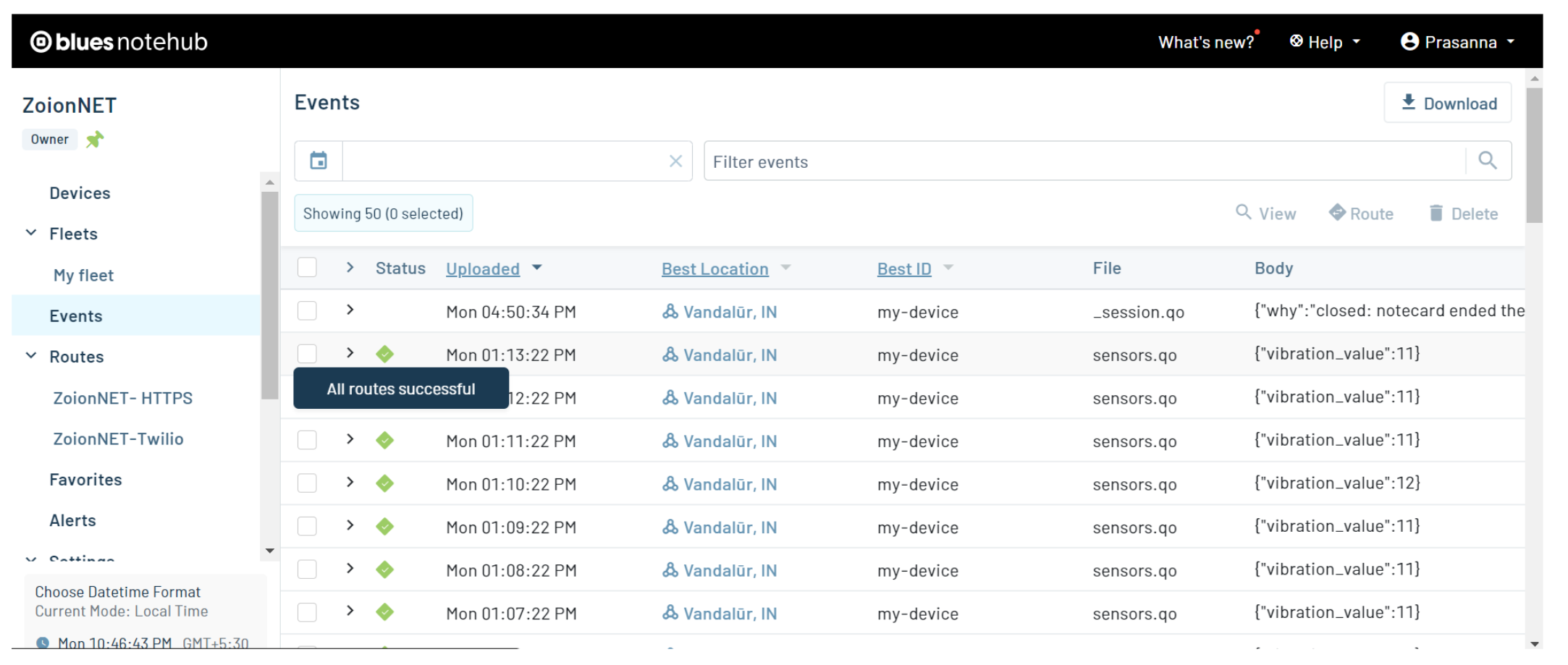}
\caption{Blues Notehub event stream from a deployed node. Successive
\texttt{sensors.qo} notes carry vibration readings; the routing status
column confirms delivery to both downstream consumers.}
\label{fig:notehub}
\end{figure}

\begin{figure}[t]
\centering
\includegraphics[width=\columnwidth]{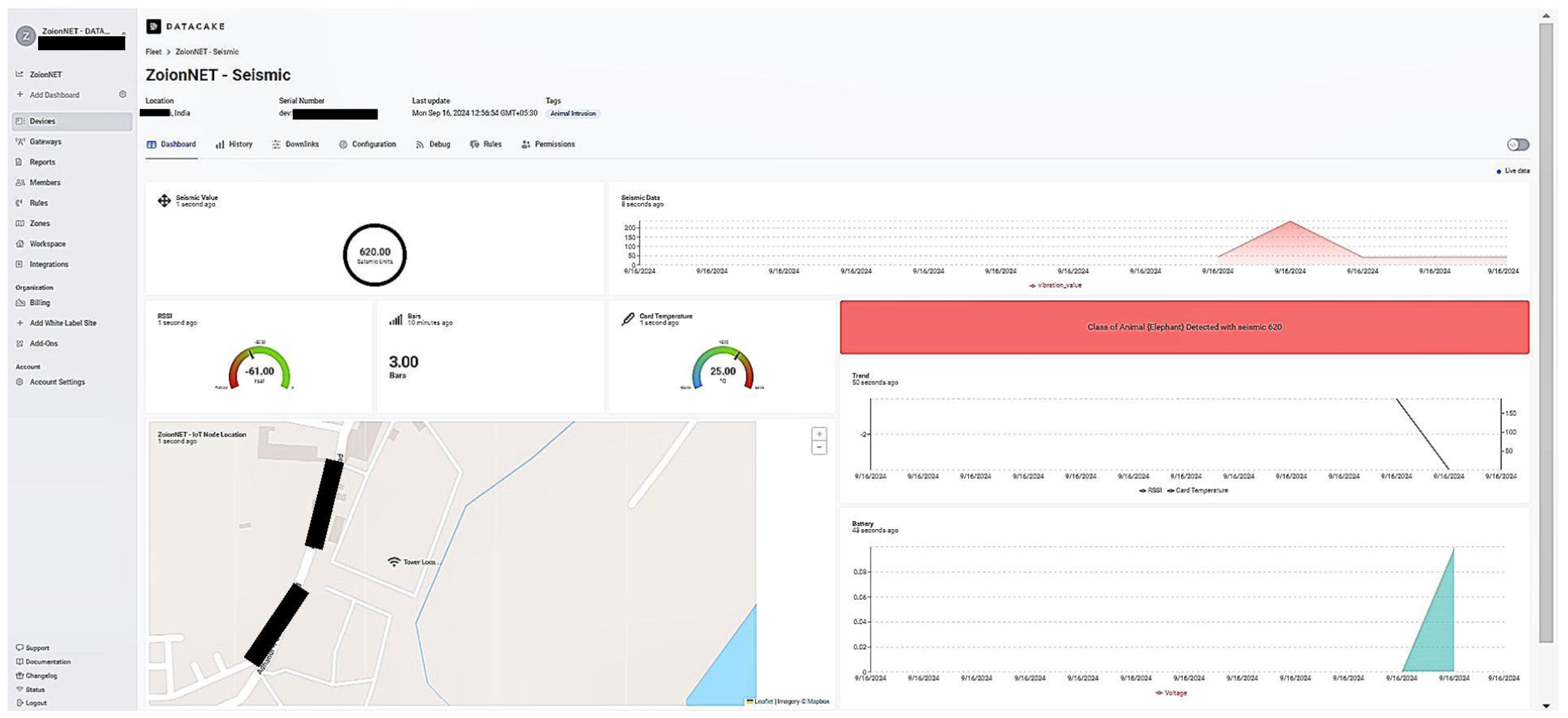}
\caption{Datacake dashboard for a deployed node, showing seismic
readings over time, current device state, node location and threshold
alarms.}
\label{fig:datacake}
\end{figure}

\begin{figure}[t]
\centering
\includegraphics[width=0.42\columnwidth]{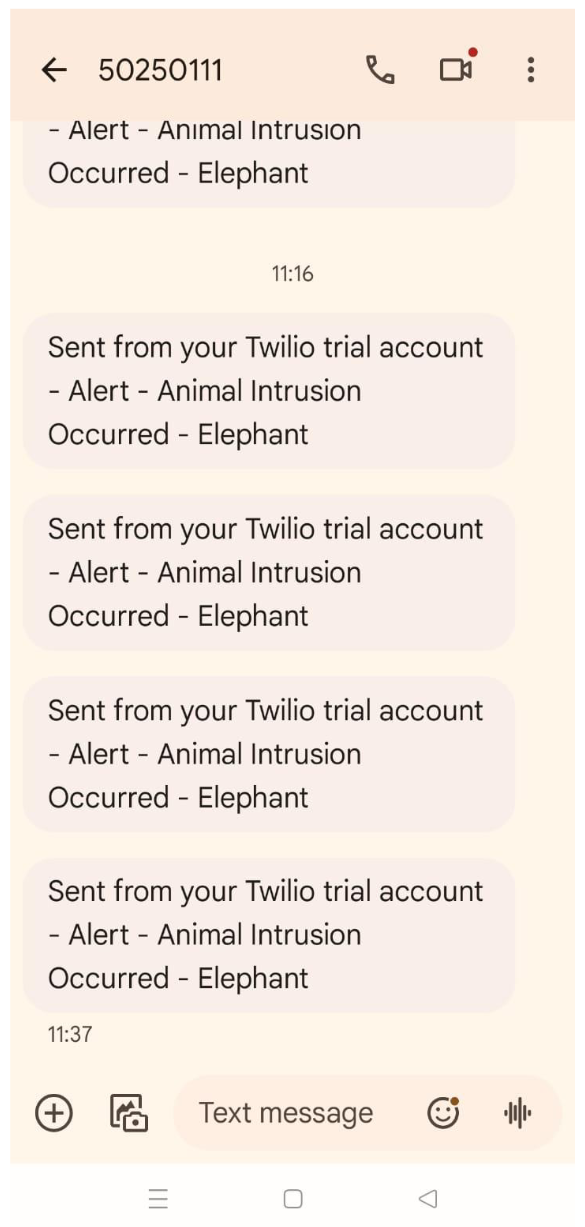}
\caption{Intrusion alerts delivered as SMS via Twilio. Each message
names the species inferred at the node.}
\label{fig:twilio}
\end{figure}

The delivery path is shown operating end to end on a deployed node:
Figure~\ref{fig:notehub} captures events arriving at Notehub,
Figure~\ref{fig:datacake} the Datacake dashboard they are routed to, and
Figure~\ref{fig:twilio} the resulting alerts on a handset. The alerts
shown were raised by the seismic channel; an acoustic detection leaves
the node through the same Notecard--Notehub route. Together these
demonstrate that the path works, and that alerts reach a person without
any local gateway infrastructure. They are not quantitative
results --- no latency, delivery-rate or false-alert measurements were
collected during the trials.

\section{Social Impact}
\label{sec:impact}

\begin{figure}[t]
\centering
\includegraphics[width=\columnwidth]{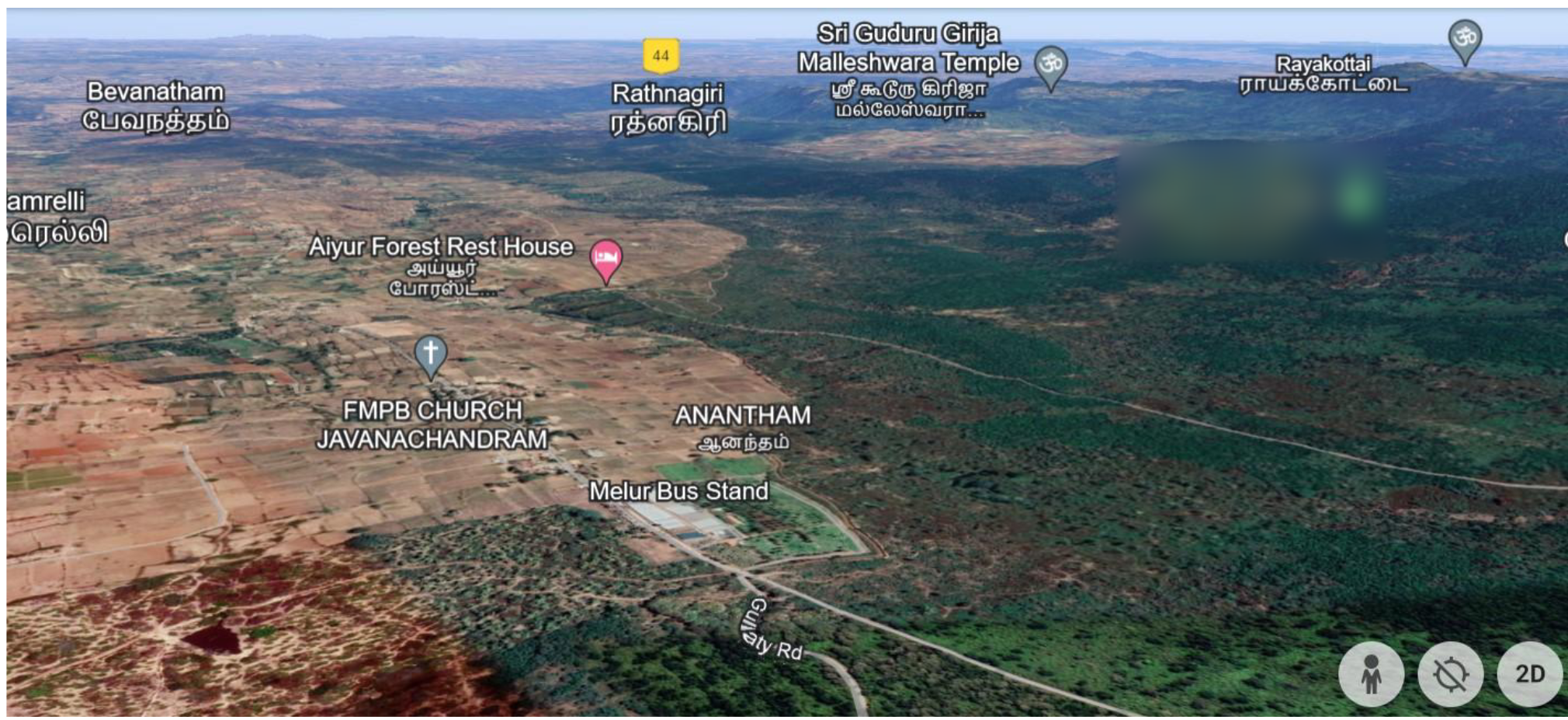}
\caption{The Melagiri landscape at the forest edge, showing the
interleaving of forest with cultivated land and settlements that produces
the conflict this system addresses. Imagery: Google Earth.}
\label{fig:map}
\end{figure}

Figure~\ref{fig:map} shows why the problem is structural rather than
incidental. Forest and cultivation interleave along the boundary,
and the settlements sit inside that interface rather than at a remove
from it. Anchetty, one of the villages along this stretch of the
Melagiri forest, alone had 16,578 residents at the 2011
Census~\cite{census2011anchetty}, and villages like it absorb the
consequences of every crossing.

For those households the immediate value of an early warning is
practical. A crop-raiding herd detected at the boundary rather than in
the field is a herd that can be deterred before the damage is done, and
without the improvised and dangerous methods --- electrified fencing
above all --- that account for a substantial share of unnatural elephant
deaths. The benefit is symmetrical: the same alert that protects a
season's income also keeps the animal away from the situations that get
it killed.

At the level of the forest department, a dashboard aggregating detections
across nodes turns individual incidents into a record. Which stretches of
boundary are crossed, at what times of year, by which species --- these
are the inputs to decisions about where to concentrate patrols, where
fencing would repay its cost, and where a corridor is being pinched.
None of that is currently visible at this site, because nothing is
recording it.

Finally, the approach generalises in a way that matters for places with
similar problems and similar budgets. Nothing in the design is specific
to the Melagiri hills: it needs cellular coverage, a species that
vocalises, and audio to fine-tune on. At the cost established in
Section~\ref{sec:results}, a boundary can be instrumented at a price
that a district forest office or a local conservation organisation can
actually meet, which is a different proposition from a system that works
only where a research budget reaches.

\section{Limitations and Future Work}
\label{sec:limitations}

\subsection{What has not been measured}

\textbf{Power.} The system is named for low power and organised around an
energy argument, and no measurement of that energy exists. No current
draw in either state, no measured battery life, no field observation of
how long a node survives on a charge. The two-state design is sound in
principle --- an idle accelerator draws less than a busy one --- but
principle is not evidence, and the central claim of this work is
currently unquantified. Measuring dormant and active current draw, and
from those deriving expected battery life at an observed duty cycle, is
the single most valuable next step and requires no new hardware.

\textbf{Delivery performance.} The alert path is shown working
end to end in Section~\ref{sec:results}, but no latency, delivery-rate or
missed-alert statistics were collected. For a warning system, time from
vocalisation to SMS is a primary metric and it is absent.

\textbf{Evaluation scope.} The reported accuracies come from held-out
audio drawn from the same public sources as the training data, not from
the deployment site. Recordings scraped from YouTube and Freesound
differ systematically from a forest boundary in India in background
noise, microphone characteristics and recording distance, so the
deployed accuracy is likely lower than the reported figures --- by an
unknown margin, since no site-recorded test set was scored.

\subsection{Scope of the system as built}

\textbf{Species coverage.} Three species are modelled. Wild boar and
gaur, both significant crop-raiders in this landscape, are not, and would
each need their own training data.

\textbf{Vocalisation dependence.} An acoustic system detects animals that
make a noise. An elephant moving quietly through vegetation produces
little that an 8~kHz microphone will catch, and the seismic channel is
the intended answer to exactly this --- but that channel is currently the
least developed part of the system, and no evaluation of it has been
recorded. Published work shows footfall detection at ranges
around 150~m is achievable~\cite{wijayaraja2025seismic}, which suggests
the modality is worth developing rather than treating as a
supplement.

\textbf{Frequency range.} Sampling at 8~kHz gives a 4~kHz Nyquist limit,
and the mel-spectrograms in Figure~\ref{fig:spectrograms} show energy
concentrated well above 512~Hz. The system therefore classifies trumpets
and similar calls rather than the infrasonic rumbles below 30~Hz that
dominate the elephant bioacoustics literature~\cite{bjorck2019elephant}
and which propagate considerably further. Rumble detection would need a
different front end, and would likely extend detection range
substantially.

\textbf{Reproducibility.} This account identifies the dataset only by its
public sources; clip counts, class balance, split sizes and per-clip
licences are not reported here, so the models can be described but not
retrained from it.

\textbf{Coverage.} Each node protects the stretch of boundary within
earshot. Instrumenting a boundary of any length means many nodes, and the
per-node cost in Table~\ref{tab:bom}, while low, multiplies.

\subsection{Future work}

The immediate priorities follow directly from the gaps above:
instrumenting a node to measure per-state current draw and field duty
cycle, building a site-recorded test set so that deployed accuracy can be separated from
benchmark accuracy, and retraining the felid class on leopard recordings
as noted in Section~\ref{sec:intro}.

Beyond that, the team intends to add a LoRaWAN
path~\cite{loraalliance2024lorawan} for locations without cellular
coverage. This matters more than it might appear: the current design is
viable at this site precisely because cellular reaches the boundary, and
the areas where conflict is worst are often the areas furthest inside the
forest, where it does not. A LoRaWAN backhaul to a gateway at the forest
edge would extend the same node design into that territory at the cost of
reintroducing the local infrastructure the cellular design was chosen to
avoid.

\section{Conclusion}
\label{sec:conclusion}

ZoionNET detects animal intrusion at a forest--village boundary by
listening rather than watching. Audio sampled at 8~kHz is converted to
mel-spectrograms on-device and classified on an Arm Ethos-U55 microNPU by
two int8-quantised YOLOv8 models, arranged so that a lightweight always-on
model gates a heavier species classifier through
exponential-moving-average thresholds. Inferences leave the node as JSON over a cellular
Notecard and arrive as a dashboard entry and an SMS, with no local
gateway anywhere in the path. The models reached reported accuracies of
0.974 and 0.946, a complete node costs \$133.57 in parts, and nodes have
been trialled in the Melagiri hills with the alert path confirmed working
end to end.

The choice of modality is the substance of the contribution. Camera-based
monitoring struggles in the farmed, inhabited landscapes where conflict
happens, and published camera-based systems addressing the same problem
run on application-class hardware. Moving to
acoustic sensing on a microNPU changes the power and cost class of the
node, which is what determines whether a boundary can actually be
instrumented at the density the problem requires.

What remains is to demonstrate that the energy argument holds in numbers
rather than in architecture. The system was designed to save power and
has never been measured doing so; establishing per-state current draw and
field duty cycle would convert the central claim of this work from a
design rationale into a result.

\section*{Acknowledgment}

The authors thank the Kenneth Anderson Nature Society (KANS) for its
support of this work, and for the local knowledge that shaped where and
how the system was deployed.

\bibliographystyle{IEEEtran}
\bibliography{references}

\begin{thebibliography}{10}
\providecommand{\url}[1]{#1}
\csname url@samestyle\endcsname
\providecommand{\newblock}{\relax}
\providecommand{\bibinfo}[2]{#2}
\providecommand{\BIBentrySTDinterwordspacing}{\spaceskip=0pt\relax}
\providecommand{\BIBentryALTinterwordstretchfactor}{4}
\providecommand{\BIBentryALTinterwordspacing}{\spaceskip=\fontdimen2\font plus
\BIBentryALTinterwordstretchfactor\fontdimen3\font minus
  \fontdimen4\font\relax}
\providecommand{\BIBforeignlanguage}[2]{{%
\expandafter\ifx\csname l@#1\endcsname\relax
\typeout{** WARNING: IEEEtran.bst: No hyphenation pattern has been}%
\typeout{** loaded for the language `#1'. Using the pattern for}%
\typeout{** the default language instead.}%
\else
\language=\csname l@#1\endcsname
\fi
#2}}
\providecommand{\BIBdecl}{\relax}
\BIBdecl

\bibitem{comsoc2024winners}
\BIBentryALTinterwordspacing
{IEEE Communications Society}, ``{IEEE ComSoc} student competition winners,''
  IEEE ComSoc website, 2024, 2024 {``Honorary Mention''} certificates.
  Accessed: 2026-11-17. [Online]. Available:
  \url{https://www.comsoc.org/engagement-community/competitions/ieee-comsoc-student-competition/winners}
\BIBentrySTDinterwordspacing

\bibitem{shaji2019kans}
\BIBentryALTinterwordspacing
K.~A. Shaji, ``How the legacy of a colonial hunter has inspired conservation
  efforts in the forests of south {India},'' Mongabay India, Jul. 2019,
  accessed: 2026-08-13. [Online]. Available:
  \url{https://india.mongabay.com/2019/07/how-the-legacy-of-kenneth-anderson-has-inspired-conservation-efforts-in-the-forests-of-south-india/}
\BIBentrySTDinterwordspacing

\bibitem{bhandari2026wire}
\BIBentryALTinterwordspacing
K.~Bhandari and R.~Tripathi, ``Why {India's} conservation science stops at the
  boundary of its protected areas,'' The Wire, Sep. 2026, accessed: 2026-11-17.
  [Online]. Available:
  \url{https://thewire.in/environment/why-indias-conservation-science-stops-at-the-boundary-of-its-protected-areas}
\BIBentrySTDinterwordspacing

\bibitem{ronoh2022hwc}
E.~K. Ronoh, S.~Mirau, and M.~A. Dida, ``Human-wildlife conflict early warning
  system using the internet of things and short message service,''
  \emph{Engineering, Technology \& Applied Science Research}, vol.~12, no.~2,
  pp. 8273--8277, 2022.

\bibitem{bjorck2019elephant}
J.~Bjorck, B.~H. Rappazzo, D.~Chen, R.~Bernstein, P.~H. Wrege, and C.~P. Gomes,
  ``Automatic detection and compression for passive acoustic monitoring of the
  {African} forest elephant,'' in \emph{Proceedings of the AAAI Conference on
  Artificial Intelligence}, vol.~33, no.~1, 2019.

\bibitem{wijayaraja2024seismic}
J.~L. Wijayaraja, J.~L. Wijekoon, M.~Wijesundara, and L.~J.~M. Wickramasinghe,
  ``Towards long range detection of elephants using seismic signals: A
  geophone-sensor interface for embedded systems,'' \emph{arXiv preprint
  arXiv:2406.05140}, 2024.

\bibitem{wijayaraja2025seismic}
J.~L. Wijayaraja, J.~L. Wijekoon, and M.~Wijesundara, ``Event detection and
  classification for long range sensing of elephants using seismic signal,''
  \emph{arXiv preprint arXiv:2509.02920}, 2025.

\bibitem{hershey2017cnn}
S.~Hershey, S.~Chaudhuri, D.~P.~W. Ellis, J.~F. Gemmeke, A.~Jansen, R.~C.
  Moore, M.~Plakal, D.~Platt, R.~A. Saurous, B.~Seybold, M.~Slaney, R.~J.
  Weiss, and K.~Wilson, ``{CNN} architectures for large-scale audio
  classification,'' in \emph{IEEE International Conference on Acoustics, Speech
  and Signal Processing (ICASSP)}, 2017, pp. 131--135.

\bibitem{gemmeke2017audioset}
J.~F. Gemmeke, D.~P.~W. Ellis, D.~Freedman, A.~Jansen, W.~Lawrence, R.~C.
  Moore, M.~Plakal, and M.~Ritter, ``Audio set: An ontology and human-labeled
  dataset for audio events,'' in \emph{IEEE International Conference on
  Acoustics, Speech and Signal Processing (ICASSP)}, 2017, pp. 776--780.

\bibitem{arm2024ethosu55}
\BIBentryALTinterwordspacing
{Arm Ltd.}, ``{Arm Ethos-U55} product brief,'' 2024, 64--512 GOP/s at 1~GHz for
  32--256 MACs per cycle; up to 90\% energy reduction for ML workloads compared
  with previous Cortex-M generations. Accessed: 2026-10-19. [Online].
  Available:
  \url{https://www.arm.com/-/media/files/pdf/product-brief/arm-ethos-u55-product-brief.pdf}
\BIBentrySTDinterwordspacing

\bibitem{jacob2018quantization}
B.~Jacob, S.~Kligys, B.~Chen, M.~Zhu, M.~Tang, A.~Howard, H.~Adam, and
  D.~Kalenichenko, ``Quantization and training of neural networks for efficient
  integer-arithmetic-only inference,'' in \emph{Proceedings of the IEEE
  Conference on Computer Vision and Pattern Recognition (CVPR)}, 2018, pp.
  2704--2713.

\bibitem{seeed2024grovevision}
\BIBentryALTinterwordspacing
{Seeed Studio}, ``Grove vision {AI} module {V2},'' Seeed Studio Wiki, {Himax}
  WiseEye2 HX6538, dual-core Arm Cortex-M55 with integrated Arm Ethos-U55,
  onboard PDM microphone. Accessed: 2026-08-13. [Online]. Available:
  \url{https://wiki.seeedstudio.com/grove_vision_ai_v2/}
\BIBentrySTDinterwordspacing

\bibitem{himax2023hx6538}
\BIBentryALTinterwordspacing
{Himax Technologies, Inc.}, ``{HX6538-A} {WE2} {AI} processor data sheet,''
  Doc.\ No.\ HX6538-A-DS, preliminary version 01, Jul. 2023, {Ethos-U55}
  microNPU with 64 MACs/cycle at up to 400~MHz. Accessed: 2026-10-19. [Online].
  Available:
  \url{https://files.seeedstudio.com/wiki/grove-vision-ai-v2/HX6538_datasheet.pdf}
\BIBentrySTDinterwordspacing

\bibitem{blues2024notecard}
\BIBentryALTinterwordspacing
{Blues Inc.}, ``Notecard cellular and notehub,'' Blues product documentation,
  cellular Notecards ship with 10 years of service and 500\,MB of prepaid data.
  Accessed: 2026-08-13. [Online]. Available:
  \url{https://blues.com/products/notecard/}
\BIBentrySTDinterwordspacing

\bibitem{datacake2024}
\BIBentryALTinterwordspacing
{Datacake GmbH}, ``Datacake {IoT} platform documentation,'' accessed:
  2026-08-13. [Online]. Available: \url{https://docs.datacake.de/}
\BIBentrySTDinterwordspacing

\bibitem{twilio2024messaging}
\BIBentryALTinterwordspacing
{Twilio Inc.}, ``Programmable messaging,'' Twilio developer documentation,
  accessed: 2026-08-13. [Online]. Available:
  \url{https://www.twilio.com/docs/messaging}
\BIBentrySTDinterwordspacing

\bibitem{allen1977stft}
J.~B. Allen and L.~R. Rabiner, ``A unified approach to short-time {Fourier}
  analysis and synthesis,'' \emph{Proceedings of the IEEE}, vol.~65, no.~11,
  pp. 1558--1564, 1977.

\bibitem{stevens1937mel}
S.~S. Stevens, J.~Volkmann, and E.~B. Newman, ``A scale for the measurement of
  the psychological magnitude pitch,'' \emph{The Journal of the Acoustical
  Society of America}, vol.~8, no.~3, pp. 185--190, 1937.

\bibitem{davis1980mfcc}
S.~B. Davis and P.~Mermelstein, ``Comparison of parametric representations for
  monosyllabic word recognition in continuously spoken sentences,'' \emph{IEEE
  Transactions on Acoustics, Speech, and Signal Processing}, vol.~28, no.~4,
  pp. 357--366, 1980.

\bibitem{font2013freesound}
F.~Font, G.~Roma, and X.~Serra, ``Freesound technical demo,'' in
  \emph{Proceedings of the 21st ACM International Conference on Multimedia},
  2013, pp. 411--412.

\bibitem{jocher2023yolov8}
\BIBentryALTinterwordspacing
G.~Jocher, A.~Chaurasia, and J.~Qiu, ``Ultralytics {YOLOv8},'' 2023, aGPL-3.0.
  Accessed: 2026-08-13. [Online]. Available:
  \url{https://github.com/ultralytics/ultralytics}
\BIBentrySTDinterwordspacing

\bibitem{redmon2016yolo}
J.~Redmon, S.~Divvala, R.~Girshick, and A.~Farhadi, ``You only look once:
  Unified, real-time object detection,'' in \emph{Proceedings of the IEEE
  Conference on Computer Vision and Pattern Recognition (CVPR)}, 2016, pp.
  779--788.

\bibitem{raspberrypi2026pi5}
\BIBentryALTinterwordspacing
{Raspberry Pi Ltd.}, ``Raspberry {Pi} 5 product brief,'' Doc.\ No.\
  RP-008348-DS, Oct. 2026, list prices: 1\,GB \$45, 2\,GB \$77.50, 4\,GB \$110,
  8\,GB \$175, 16\,GB \$305. Accessed: 2026-10-19. [Online]. Available:
  \url{https://pip.raspberrypi.com/documents/RP-008348-DS-raspberry-pi-5-product-brief.pdf}
\BIBentrySTDinterwordspacing

\bibitem{census2011anchetty}
\BIBentryALTinterwordspacing
{Census2011.co.in}, ``{Anchetty} village population -- {Denkanikottai} --
  {Krishnagiri}, {Tamil Nadu},'' Population Census 2011 data, accessed:
  2026-11-19. [Online]. Available:
  \url{https://www.census2011.co.in/data/village/644087-anchetty-tamil-nadu.html}
\BIBentrySTDinterwordspacing

\bibitem{loraalliance2024lorawan}
\BIBentryALTinterwordspacing
{LoRa Alliance}, ``{LoRaWAN} specification,'' {Low} Power Wide Area networking
  specification. Accessed: 2026-08-13. [Online]. Available:
  \url{https://lora-alliance.org/about-lorawan/}
\BIBentrySTDinterwordspacing

\end{thebibliography}

\end{document}